\documentclass[intlimits,twoside,a4paper]{article}

\usepackage[cp1251]{inputenc}
\usepackage{multirow}
\usepackage[eqsecnum]{cmpj3}

\usepackage{bm}
\articletype{Rapid communication}

\issue{2022}{25}{4}{44201}
\doinumber{10.5488/CMP.25.44201}

\title[Water-DMSO liquid mixtures]
{Apparent molar volume anomaly in
water-dimethyl sulfoxide liquid mixtures.
Molecular dynamics computer simulations}

\author[M. Aguilar, H. Dominguez, O. Pizio]
{M. Aguilar\orcid{0000-0003-3850-1188}\refaddr{label1},
H.Dominguez\orcid{0000-0001-6126-9300}\refaddr{label2},
        O. Pizio\orcid{0000-0001-8333-4652}\refaddr{label1}
\thanks{Corresponding author: \email{oapizio@gmail.com}.}}

\addresses{
\addr{label1}Instituto de Qu\'{i}mica, Universidad Nacional Aut\'{o}noma de M\'{e}xico,
Circuito Exterior, 04510, Cd. Mx., M\'{e}xico
\addr{label2} Instituto de Investigaciones en Materiales, Universidad Nacional Aut\'{o}noma de M\'{e}xico,
Circuito Exterior, 04510, Cd. Mx., M\'{e}xico}


\date{Received August 9, 2022, in final form October 19, 2022}
\begin{document}

\maketitle

%%%%%%%%%%%%%%%%%
\begin{abstract}
We have studied the composition dependence of density
of liquid water-DMSO mixtures at different temperatures
by using the isobaric-isothermal (NPT) molecular dynamics computer simulations. The
non-polarizable semi-flexible, P1 and P2 models for the DMSO molecule
combined with the TIP4P-2005 water
model are considered. 
%We restrict calculations to atmospheric pressure, 0.1013 MPa, and to 
%temperatures: 298.15K, 318.15K and 338.15K.
The excess mixing volume and the apparent molar volumes of the species are reported.
We have established that the P1-TIP4P-2005 model for the mixture
provides a very good description of the location of the minimum of
apparent molar volume for DMSO species indicating the anomaly. Most important is that
the  temperature interval where the hydrophobic effect exists,
is correctly captured with this modelling, in contrast to the P2-TIP4P-2005 model.

\printkeywords
%
%
%\pacs 02.70.Ns,61.20.Ja,82.30.Rs,87.15.hp
\end{abstract}

Liquid mixtures of dimethyl sulfoxide (DMSO) and water are widely used in organic chemistry, 
chemical engineering, and biomedical applications. 
In this rapid communication we refer to the static light scattering experiment
performed on aqueous solutions of dimethyl sulfoxide (DMSO) at various compositions and 
at two temperatures, 20$^{\circ}$C and 50$^{\circ}$C ~\cite{rodnikova}.  
In the concentration range around 10 mol percent of
DMSO, an abnormal maximum of scattered light was detected, 
the intensity of which decreases with an increase of temperature.
This maximum has been attributed to the hydrophobic effect and has been interpreted in
terms of the minimum of apparent molar volume of DMSO upon concentration.
Inspired by this report, we have undertaken the study of
the  composition dependence of density of water-DMSO liquid mixtures at
298.15~K, 318.15~K and 338.15~K by using molecular dynamics computer simulations
complementing our recent study performed at 298.15~K~\cite{aguilar}.
Our focus is on the capability of simple non-polarizable models for DMSO and water 
to reproduce the experimental observations by capturing peculiar composition interval
in the restricted temperature range.  

Previously, binary mixtures of water with DMSO, in particular at a low DMSO concentration, 
were the subject of study in different 
laboratories~\cite{bagchi1,bagchi2,mancera1,mancera2,gujt,aguilar}. 
Specifically, in \cite{bagchi1,bagchi2} the focus is on the composition trends
of certain properties at a single temperature 300~K. The NPT setup was used.
On the other hand, the NVT simulations at 298.15~K, 318.15~K and 338.15~K and at
a fixed DMSO molar fraction ($X_{\text{dmso}}=0.055$) were performed. The experimental 
density values were assumed.  In this study, we have chosen quite popular, nonpolarizable, 
P1 and P2 models for DMSO species~\cite{luzar}, from a huge set of force fields
for DMSO~\cite{chalaris,idrissi},   
and well tested TIP4P-2005 water model~\cite{vega}. 
Technical details of our isobaric-isothermal molecular dynamics computer simulations
were described very recently in the previous publication on the topic~\cite{aguilar}.

\begin{figure}[!t]
\begin{center}
\includegraphics[width=6cm,clip]{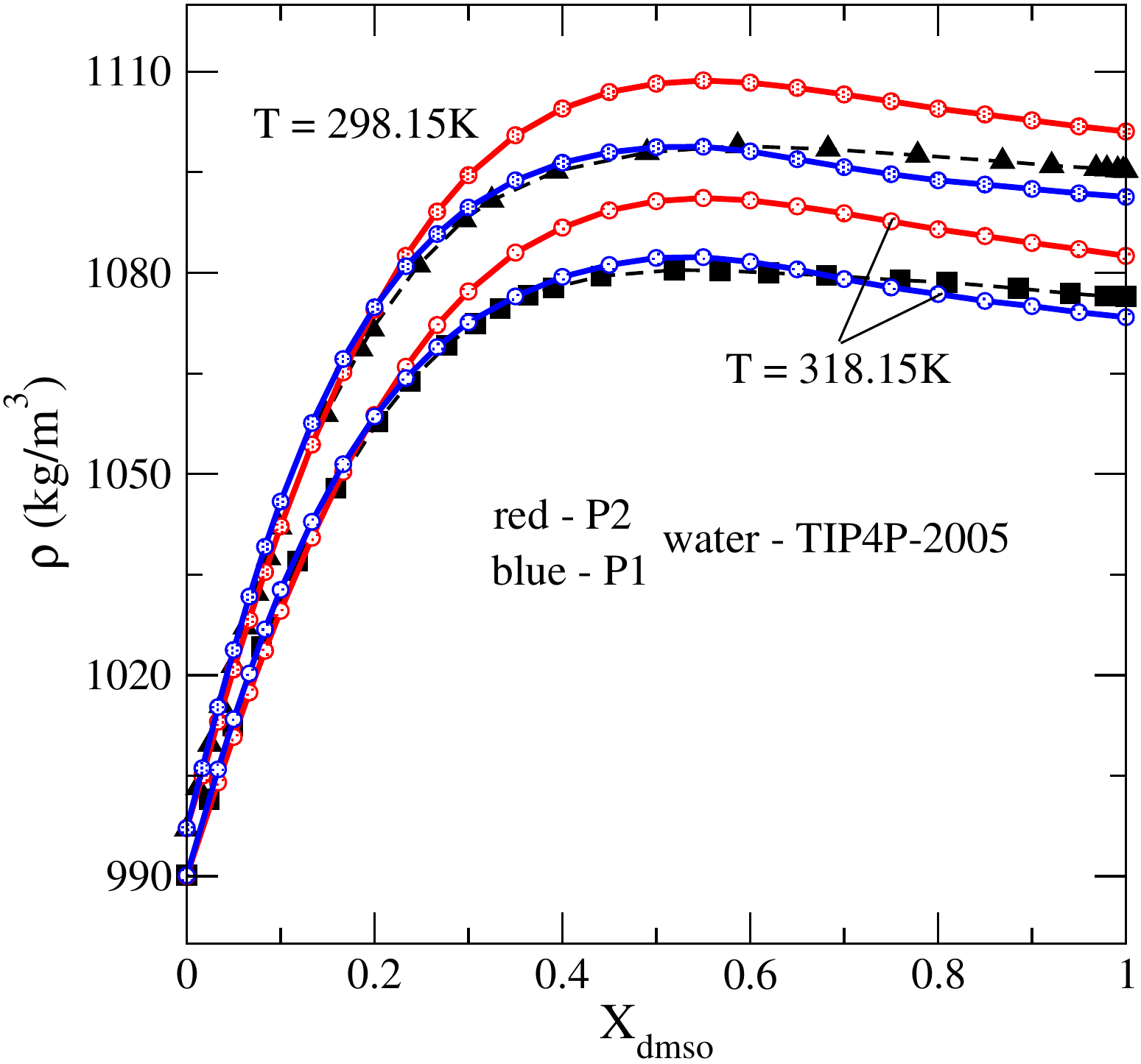}
\includegraphics[width=6cm,clip]{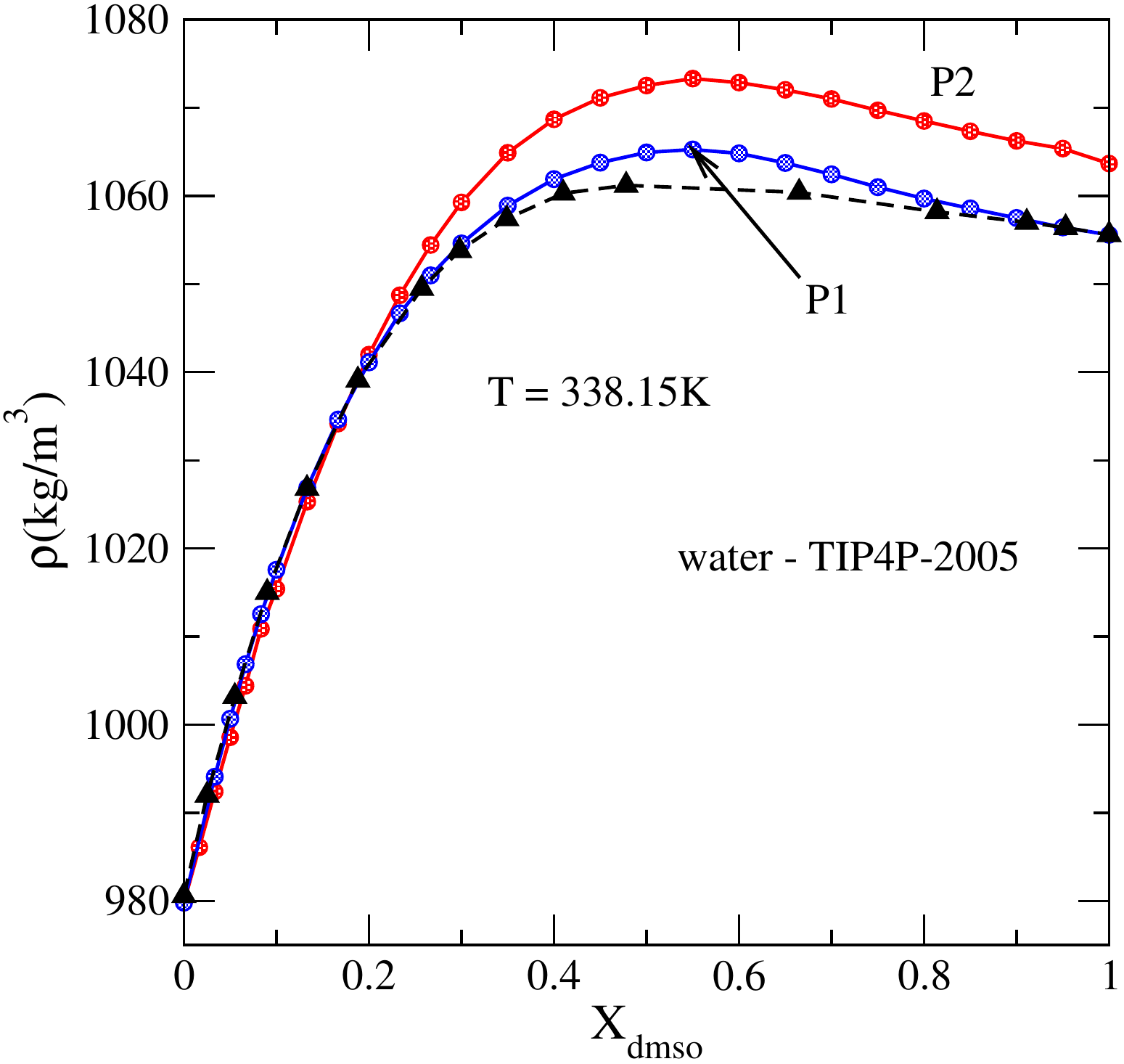}
\end{center}
\caption{(Colour online) Panel a: Composition dependence of the density of water-DMSO
mixtures from the NPT MD simulations of  P2-TIP4P-2005 model (solid red lines),
P1-TIP4P-2005 model (solid blue lines)
in comparison  with the experimental data (dashed lines with triangles) at 298.15~K~\cite{egorov}
and at 318.15~K (dashed line with squares) \cite{carmen}.
Panel b: Similar to the panel a, but at 338.15~K. Experimental data (dashed line with triangles) 
are from~\cite{cowie}.
}
\label{fig1}
%\protect
\end{figure}

As common, the strategy of exploration is to describe a desired property
on composition and to capture its deviation from ideality as well. We do that for
the mixture density at each fixed temperature. A set of results is shown in figure~\ref{fig1}.
The panel a of this figure refers to $T = 298.15$~K and $T = 318.15$~K, whereas the
panel b illustrates the results at $T = 338.15$~K. This temperature was also  chosen
in~\cite{mancera1} due to the availability of experimental data.
From the inspection of the results, we observe that the P1-TIP4P-2005 and P2-TIP4P-2005
are quite accurate in the interval of composition from pure water, $X_{\text{dmso}} = 0$, up to
$X_{\text{dmso}} \approx 0.2$. At a higher DMSO content in the mixture, the simulation data
begin to deviate from the experimental results. The maximum density at a certain 
composition, $X_{\text{dmso}} \approx 0.58$, is captured by both models in question.
Nevertheless, it can be seen that
the P1-TIP4P-2005 model performs much better than its P2-TIP4P-2005 counterpart.
This latter model overestimates the density in the interval above $X_{\text{dmso}} >$ 0.2.
Moreover, the deviation from the experimental points does not decrease while the temperature
grows from 298.15~K up to 338.15~K. The P1-TIP4P-2005 model exhibits a small inaccuracy for density
at ``intermediate'' compositions, but performs much better than P2-TIP4P-2005.
It is difficult to judge the precision of computer simulations data at low $X_{\text{dmso}}$ values
at the scale in the figure~\ref{fig1}. Better insights follow from the excess properties.  

We describe geometric aspects of mixing of the species in terms of
the excess mixing volume defined as common, $\Delta V_{\text{mix}} = V_{\text{mix}} - X_d V_d - (1-X_d) V_w$,
where $V_{\text{mix}}$, $V_d$ and $V_w$ refer to the molar volume of the mixture and
of the individual components, DMSO and water, respectively.
Apparently, this property is not strongly dependent on temperature in the interval we deal with.
This follows from the experimental results shown in figure~\ref{fig2}. The mixing volume slightly decreases 
in magnitude upon increasing temperature, as expected. In general terms, computer simulation results
show that the excess mixing volume is overestimated in the framework of models assumed. Thus,
the mixture of water and DMSO species from simulations is predicted to be more non-ideal 
than its laboratory counterpart in almost entire composition range at $T =$ 298.15~K. 
The $\Delta V_{\text{mix}}$ values from simulations decrease in magnitude in agreement with experimental trends.
However, the location of the minimum along $X_{\text{dmso}}$ axis from simulations and experiments 
is slightly different. Finally, one can observe that the performance of the P1-TIP4P-2005
model is slightly better compared to the P2-TIP4P-2005 model.

\begin{figure}[!t]
\begin{center}
\includegraphics[width=6cm,clip]{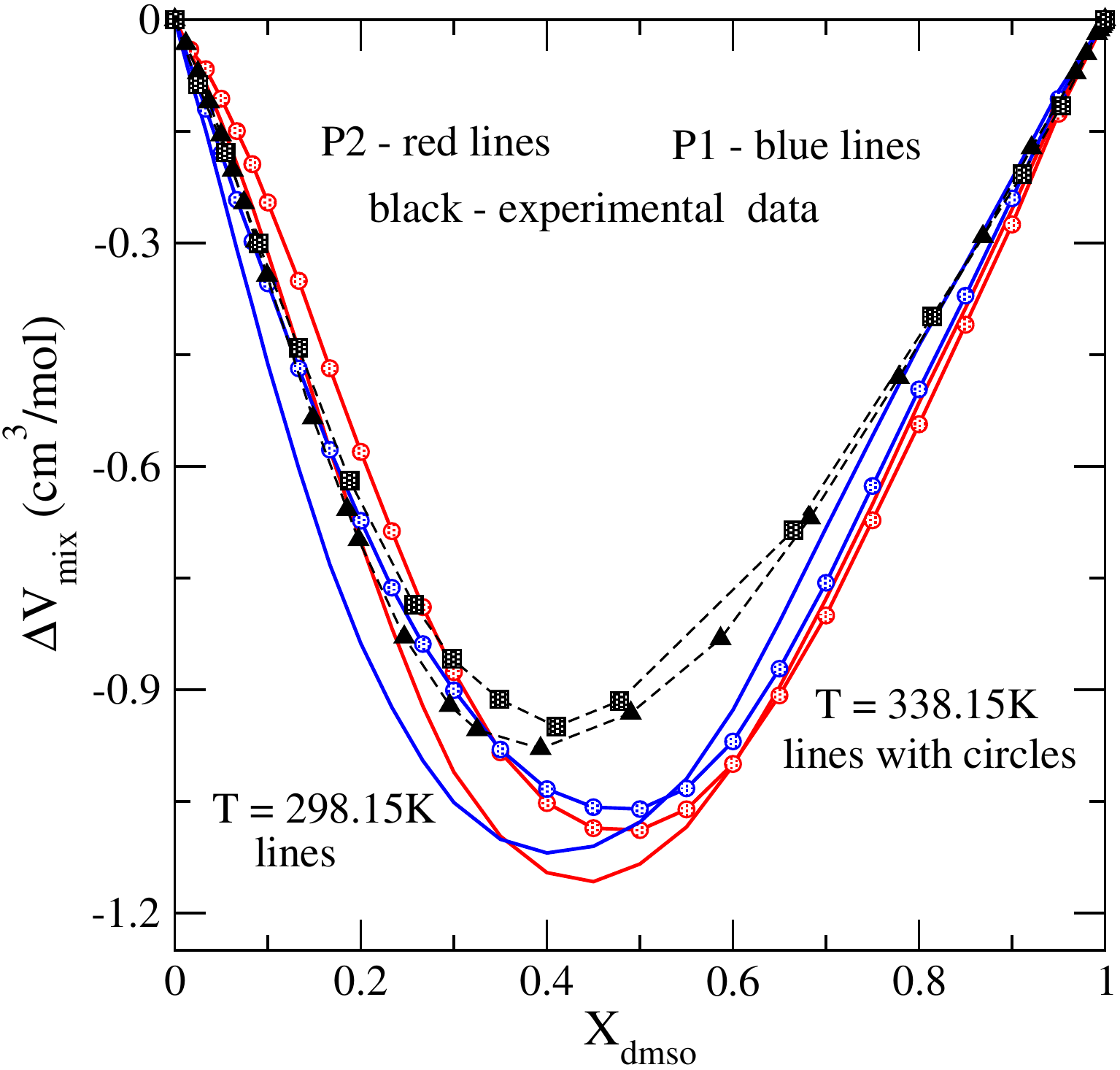}
\end{center}
\caption{(Colour online) A comparison of the composition dependence of the excess mixing volume
of water-DMSO mixtures for two DMSO models, P1 and P2, combined with TIP4P-2005 water
model,  with
the experimental data at 298.15~K (\cite{egorov}---black
triangles) and at 338.15K (\cite{cowie}---black squares).
}
\label{fig2}
\end{figure}

In order to discern the contribution of each species of the mixture into
the mixing volume, it is common to resort to the excess partial molar volumes as we have done
recently~\cite{aguilar}.
However, similar insights into the geometric aspects of mixing on composition,
both from experiments and simulations, can be obtained by resorting to the notion of
the apparent molar volume of species, rather than to the partial molar volumes.

\begin{figure}[!t]
\begin{center}
\includegraphics[width=6cm,clip]{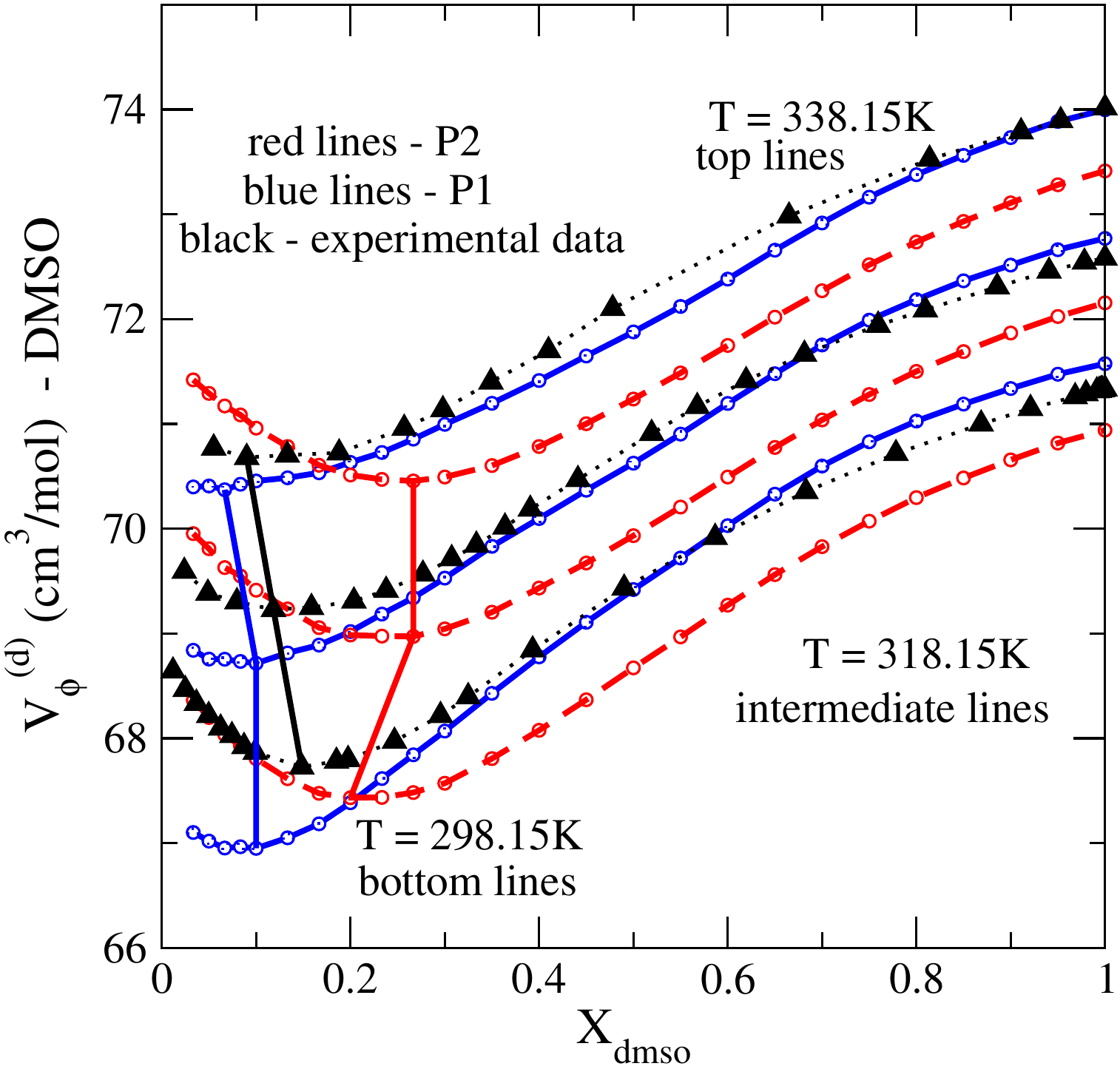}
\end{center}
\caption{(Colour online) 
A comparison of the composition dependence of the
apparent molar volume of DMSO from simulations
with the experimental data marked as triangles from~\cite{egorov} at
298.15~K, from~\cite{carmen} at 318.15~K and from \cite{cowie} at 338.15~K.
}
\label{fig3}
\end{figure}

The apparent molar volume
for each species, according to the definition~\cite{torres}, is:
$V_{\phi}^{(w)}= V_w + \Delta V_{\text{mix}}/(1-X_d)$ and  $V_{\phi}^{(d)}= V_d + \Delta V_{\text{mix}}/X_d$.
We elaborated the experimental density data from~\cite{egorov,carmen}
and the results from our simulations to construct the plots shown in  figures~\ref{fig3} and \ref{fig4}.
The most important and illuminating results for the apparent molar volume for
DMSO species, $V_{\phi}^{(d)}$, are shown in figure~\ref{fig3}. Namely, we observe that both models in
question predict the existence of miminum at certain composition $X_{\text{dmso}}$. The
points describing the location of the minimum at different temperatures from simulations
and experiments are joined by respective lines. Evolution of the minimum with
temperature is appropriately captured by the P1-TIP4P-2005 model in contrast
to the P2-TIP4P-2005 model. Moreover, the P1-TIP4P-2005 model correctly predicts that
the hydrophobic trends diminish with increasing temperature. Performance of the
P2-TIP4P-2005 model in this aspect contradicts the experimental trends 
and is not satisfactory. We have obtained the minimum
for $V_{\phi}^{(d)}$ for this model,  even at a higher temperature equal 
to 358.15~K (the respective plot is not shown for economy of space).

Concerning the behaviour of the apparent molar volume of water species, $V_{\phi}^{(w)}$,
upon composition (figure~\ref{fig4}), it is worth to mention a rather satisfatory performance of the
P1-TIP4P-2005 model in the entire composition range, especially for higher  temperatures.
The respective figure at 298.15~K is given in~\cite{aguilar} (figure~3d).
It is necessary to mention, however, that the experimental results for DMSO-rich 
(more specifically, extremely rich) mixtures
are difficult to obtain precisely, because high-purity DMSO is an exceedingly
hygroscopic solvent, the discussion of this issue is given in \cite{egorov}.

To conclude, we have established that the P1 and P2 models for DMSO combined with the
TIP4P-2005 water model yield a minimum of the apparent molar volume for DMSO
species at low values for $X_{\text{dmso}}$. This behavior is the manifestation of 
hydrophobic effects in these mixtures at a specific composition interval. However,
it appears that the predictions from  P1-TIP4P-2005 model agree better with the
experimental trends  than the ones from P2-TIP4P-2005 model. Moreover,
the P1-TIP4P-2005 model predicts the existence of a minimum of $V_{\phi}^{(d)}$ for
DMSO within the correct temperature interval from 298.15~K up to 338.15~K. 
At this highest temperature,
the minimum almost disappears, as deduced from experimental data. 
By contrast, P2-TIP4P-2005 model does not 
provide an accurate estimate for a peculiar composition interval and does not
yield a decay for the hydrophobicity trends on temperature. 
In various publications this model is claimed to be the best, see for
example \cite{kovalenko}. Here, we show that this conclusion is
valid in certain aspects at a room temperature only. In addition, it is important to
mention that the combination of the OPLS model for DMSO with TIP4P-2005 model 
for water does not provide a correct composition dependence of density
at room temperature, cf. figure~2 of~\cite{aguilar} and at temperatures 
of this study. In consequence, the anomaly of the apparent molar volume
is not captured by this type of model at all. These results are available upon request.
 
\begin{figure}[h]
\begin{center}
\includegraphics[width=6cm,clip]{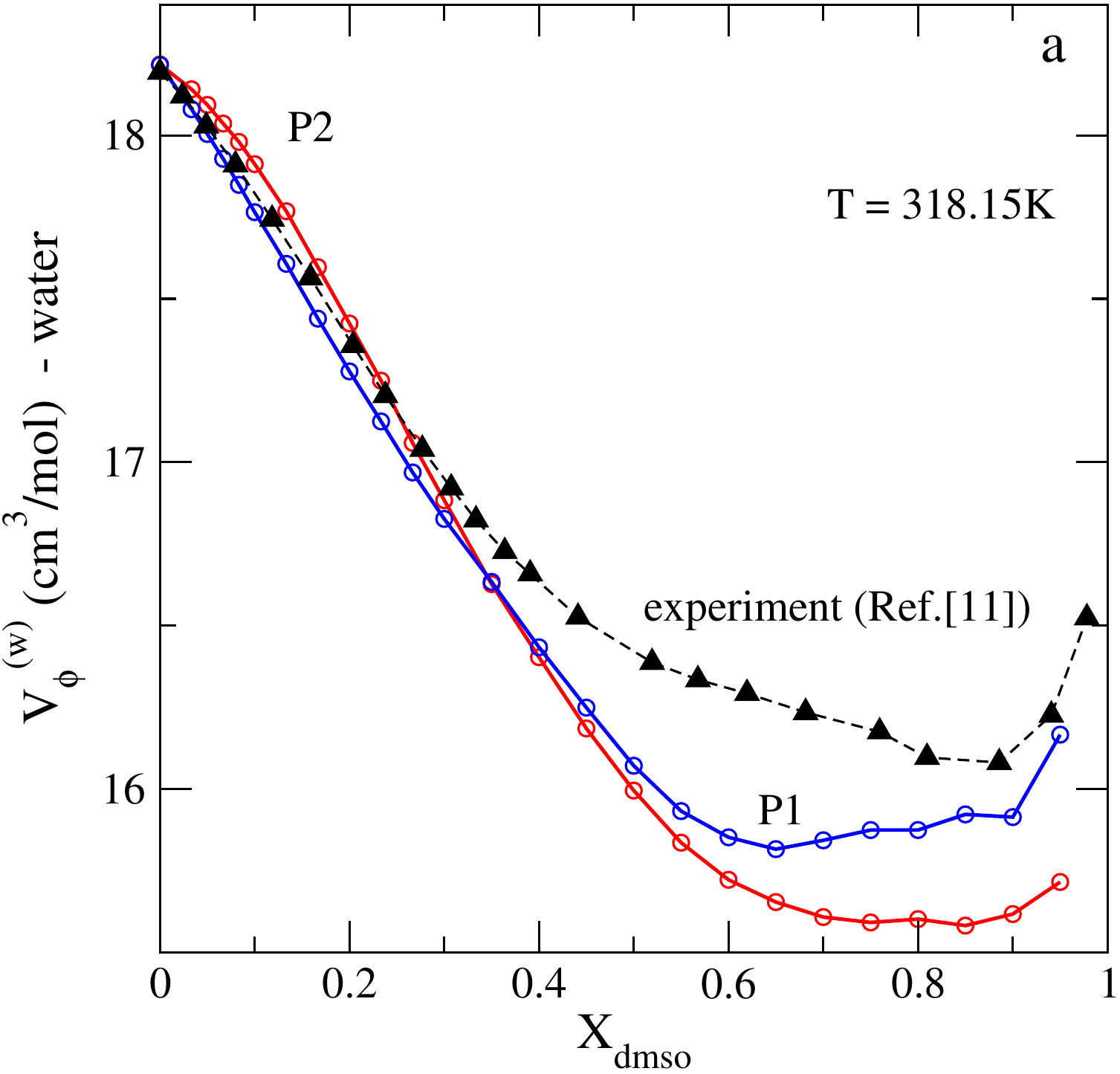}
\includegraphics[width=6cm,clip]{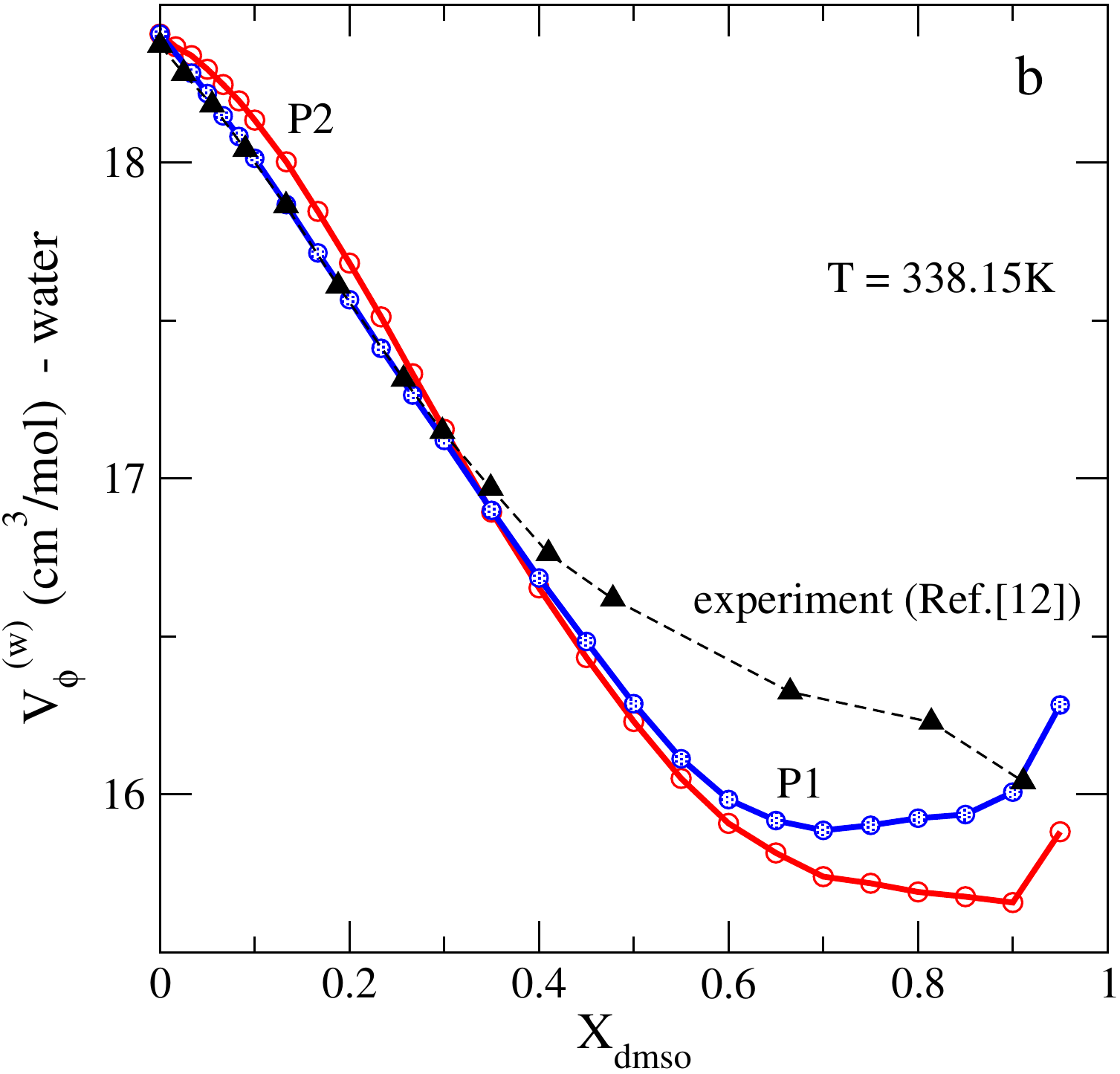}
\end{center}
\caption{(Colour online) 
Panels a and b: A comparison of the composition dependence of the
apparent molar volume of water from simulations
with the experimental data (triangles \cite{carmen}) at 318.15~K
and \cite{cowie} at 338.15~K. Similar plot at 298.15~K is shown
as figure~3d of~\cite{aguilar}.
}
\label{fig4}
\end{figure}

Previously, we  explored the minimum of the methanol apparent molar volume on composition 
for water-methanol mixtures~\cite{mario} at 298.15~K, see also~\cite{chechko} for
the interpretation of the observed phenomena. This kind of system is the mixture of 
two hydrogen bonded liquids. By contrast, in the systems of the present study there is no bonding between
the DMSO solutes in water. Apparently, the temperature trends should be slightly different
for these two classes of systems. 

In addition, it seems intriguing to investigate the evolution of 
trends of hydrophobicity in water-DMSO mixtures upon changing the external pressure as well. 
One can be guided by important experimental observations, see e.g., \cite{kanno,taschin}. 
More generally, from the simulations perspective
it is challenging to construct a more complete map of anomalies, in pressure, temperature
and composition variables, that would involve, for example, the temperature of maximum density and 
the minimum of isothermal compressibility, even by using the non-polarizable models.
These data would serve as a useful benchmark for the following development of polarizable models.
In the case of water-DMSO mixtures, certain progress has been reached at this level of 
modelling in~\cite{bachmann1,bachmann2}. However, the BK3 polarizable water model
seems to be a more convenient starting point,  because the principal anomalies of pure 
water are quite well captured in this framework~\cite{andras1,andras2}.  
Some of these issues are under study in our laboratory.

\newpage
\ukrainianpart

\title
{Аномалія видимого молярного об’єму рідких сумiшей вода-диметилсульфоксид. Комп'ютерне моделювання методом молекулярної динамiки%
}
\author
{M. Агілар \orcid{0000-0003-3850-1188}\refaddr{label1},
	Е. Домінгес\orcid{0000-0001-6126-9300}\refaddr{label2},
	O. Пізіо\orcid{0000-0001-8333-4652}\refaddr{label1}}

\addresses{
	\addr{label1}
	Інститут хімії, Національний автономний університет Мехіко, 
	Circuito Exterior, 04510, Мехіко, Мексика
	\addr{label2} Інститут матеріалознавства, Національний автономний університет Мехіко,
	Circuito Exterior, 04510, Мехіко, Meксика}

\makeukrtitle

\begin{abstract}
	Досліджено концентраційну залежність густини рідких сумішей вода-диметилсульфоксид
	(ДМСО) при різних температурах з використанням комп’ютерного моделювання методом iзобарично-iзотермiчної молекулярної динамiки. 
	Розглядаються неполяризовні напівгнучкі моделі  P1 і P2 для молекули ДМСО
	у поєднанні з моделлю  TIP4P-2005 для води. 
	Наведено результати обчислень для надлишкового об’єму змiшування та видимих молярних об’ємiв компонент суміші.
	Встановлено, що модель P1-TIP4P-2005 для цієї суміші забезпечує дуже добрий опис положення мінімуму
	видимого молярного об’єму компоненти ДМСО, який вказує на аномалію. Найбільш важливим є те, що вибрана модель,
	на відміну від її відповідника P2-TIP4P-2005, правильно передбачає наявність температурного інтервалу, в якому 
	існує гідрофобний ефект.
\end{abstract}

\end{document}